# Propose a Fuzzy Queuing Maximal Benefit Location Problem

*Department of Industrial and Mechanical Engineering, Qazvin Islamic Azad University, Qazvin, Iran*

**Abstract**

This paper presents a fuzzy queuing location model for congested system. In a queuing system there are different criteria that are not constant such as service rate, service rate demand, queue length, the occupancy probability of a service center and Probability of joining the queue line. In this paper with fuzzifying all of these variables, will try to reach an accurate real problem. Finally we change the problem to a single objective function and as far as this model is in NP-Hard classification we will use genetic algorithm for solving it and ant colony for comparison is used for their results and run time.

## 1. Introduction

Determination of where to locate servers and how many servers to have in a given area is perhaps the most important decision faced by any Company or organization. The field of set covering is a practical concept with a vast usage in Air port hubs [1], Blood bank [2], Emergency Medical Services [3], fast food restaurants [4], Fire stations [5], Telecommunication switching centers [6], Location of bank accounts [7] and vehicle routing [8] and a lot of more usage was found for this problem. Much research has been carried out on location problem in which it is required to minimize total travel time, physical distance, or some other travel related "Cost", and it is often assumed that facilities are sufficiently large to meet any demand likely to be encountered [9]. Location of service facilities and allocation of service calls to servers, dramatically are being affected by the congestion of the demand. All of the models are designed based on the providing the highest level of the service and achieving the lowest level of congestion possible.

Current et al. introduced eight basic facility location models, which are set covering, maximal covering, p-center, *p*-dispersion, *p*-median, fixed charged, hub and maximum. In all of them, the general problem is to locate new facilities to optimize distance or some measures more or less functionally related to distance (e.g. travel time or cost, demand satisfaction). The first four are based on maximum distance and the second four are based on total (or average) distance [10].

As far as our model is in maximal covering models category a brief review of these model will be beneficial. The location set covering problem (LSCP) as a version of set covering problem was introduced by Toregas *et al.* in 1971 [11]. The next step in this field is introducing Maximal

Covering Location Problem (MCLP) by church and ReVelle in 1974 [12]. The idea of Server congestion was first considered by Larson. Before 1983 all models that are introduced are not probabilistic models, but Daskin in 1983 [13] built the structure of probabilistic models with MEXCLP that is probabilistic version of MCLP. Later, Berman et al. [14-16] developed some models using queuing theory for congested networks. Then Marianov *et al.* [17-19] proposed several models in which the number of requests for service was stochastic process.

Real situations very often have demand for their services which is both variable and random in nature. Then, although the facility may be able to cope with average demand, there will be times of heavy demand when it will not cope; such a facility will be said to be congested. For congested systems, a facility will not be able to cope at times of heavy demand. When this is the case, it will possible for users to wait until the facility is free to serve them whereas in some other cases such as, for example maternity homes, it is not feasible to wait. When waiting is not permitted (or only limited waiting is allowed) then a user is lost to fully occupied facility and their demand is either demand is either satisfied elsewhere or not at all. A natural objective to minimize in this case will thus be the total amount of demand lost to the system [9].

The conditions based on the knowledge for the users resulted in formulated the congested location problem as follows:

First, when users have very little knowledge of queue characteristics. Second, when users have estimates of mean queue length for all facilities . Third, when users have knowledge of current state of relevant queues.

A location problem involves users traveling to a facility for service, or server traveling for facilities to the users [9]. We will consider immobile (fixed) servers in this paper.

Although stochastic models can cater for a variety of cases, they are not sufficient to describe many other situations, where the probability distribution of customer's demands may be unknown or partially known. For example, we want to establish some manufacturing factories in new regions to service some new customers whose demands can neither be given precisely nor from history data. But those demands can be described by the natural language such as large, little or general, etc. In these cases, fuzzy set theory may do better in dealing with ambiguous information. Fuzzy set theory was initialized by Zadeh [20] and has been widely applied in many real problems. It has been proved to be a useful tool to solve problems with uncertainty [21]**.** Humans are unsuccessful in making quantitative predictions, whereas they are comparatively efficient in qualitative forecasting. Further, humans are more prone to interference from biasing tendencies if they are forced to provide numerical estimates since the elicitation of numerical estimates forces an individual to operate in a mode which requires more mental effort than that required for less precise verbal statements [22]. In the facility location problem, the conventional approaches tend to be less effective in dealing with the imprecision or vagueness nature of the linguistic assessment. There has been an increasing interest for fuzzy sets to be used for the facility location problem in the recent years [23].

In the past decades, there are many people who have brought fuzzy theory into facility location problem. For example, in Bhattacharya *et al*. [24, 25] new facilities are considered to be located under multiple fuzzy criteria, and a fuzzy goal programming approach has been developed to deal with the problems. In Cano´s *et al.* [26], a fuzzy set of constraints is introduced into the classical p-median problem, And the decision is made which provides significantly lower costs by leaving a part of the demand uncovered.

Also Chen and Wei [27], Darzentas [28], Rao and Saraswati[29] have discussed various facility location problems by fuzzy logic methods. However, all the parameters in these problems are deterministic, and fuzzy theory is only used to solve the classical mathematical programming effectively. Zhou and Liu [30] assumed that the demands of customers are fuzzy variables, and will give some new fuzzy programming models for Location problem. Zhou [31] assumed that locations of customers to be fuzzy and some fuzzy programming models are proposed for Minimax Location Problem (MLP) under the minmax criterion.

Real situations very often have demand for their services which is both variable and random in nature. Then, although the facility may be able to cope with average demand, there will be times of heavy demand when it will not cope; such a facility will be said to be congested. For congested systems, a facility will not be able to cope at times of heavy demand. When this is the case, it will possible for users to wait until the facility is free to serve them whereas in some other cases such as, for example maternity homes, it is not feasible to wait. When waiting is not permitted (or only limited waiting is allowed) then a user is lost to fully occupied facility and their demand is either demand is either satisfied elsewhere or not at all. A natural objective to minimize in this case will thus be the total amount of demand lost to the system [9].

In this paper we will consider some servers location in some nodes with queuing theory. In a queuing system there are different criteria that are not constant such as service rate, demand rate, queue length, the occupancy probability of a service center and Probability of joining the queue line that are change based on the servicing time and the market that this queue is located in it. Considering a constant value for each of these variables will cause that the problem will not be a realistic problem. The conditions are vague and changeable, so the probabilistic and fuzzy solutions are proposed to overcome this weakness. As for a probabilistic approach, one can fit probability distributions on the basis of the stochastic experiments and the recorded data. This approach leads to the estimation of the model's parameters and ultimately the structure of the model. As for a fuzzy approach, In situations where there are no reliable recorded data for estimation purposes, we can estimate the parameters imprecisely on the basis of our perceptions. In fact, instead of gathering data for statistical estimation of parameters by spending time and cost, one can develop and analyze the model on the basis of the imprecise data. We choose to use fuzzy approach to overcome this weakness. A novel Maximal covering Location Problem with fuzzy will be proposed and genetic and Ant colony Optimization will be compared in this model to find out which of them are more suitable for solving the model in this paper

The content of this paper is outlined as follows. In section 2, Motivation of applying fuzzy theory and the basic definition is given. The problem formulation is considered in sections 3 .mathematical modality is given in section 4.In section 5 The Fuzzy Queuing Maximal Benefit Location Model (FQMBLM) is proposed. Solution algorithms including genetic algorithm and Ant Colony Optimization algorithm are presented in section6. Some numerical problems are solved based on the algorithms and the results are given in Section 7.In Section 8 the conclusions and further research is considered.

## 2. Preliminaries

For the better understanding of this paper , let us first review the concepts of fuzzy set, convex, normal and the introduce the concept of fuzzy number, membership function and triangular fuzzy numbers.

**Definition 1**. If X is a collection of objects denoted generically by x, then a fuzzy set in of X is a set ordered pairs [32]:

$$ = \left\{ (x\,,\mu\ \ (x))\ |x \in X \right\}$$

Where the symbol x denotes the element of the set X and μ (x) is called the ship function or member the degree of membership function or the degree of membership of x in that maps X to the membership space [0,1].

**Definition 2**. A fuzzy set is convex if [32]:

$$\mu\ \ (\lambda\ \ _{1}+(1-\lambda)\ \ _{2}) \geq \min\{\mu\ \ (\ \ _{1}),\mu\ \ (\ \ _{2})\},\qquad _{2},\ \ _{1}\epsilon X\,,\ \lambda\epsilon[0,1]$$

Definition 3.if μ (x)=1,the fuzzy set is called normal.[32]

Definition 4.A fuzzy number is a convex normalized fuzzy set . [32]

Definition 5.the membership function μ (x) of intersection = ∩ is point wise defined by

$\mu$ ( ) = {$\mu$ ( ), $\mu$ (x)} , x∊ X

OR

$\mu$ = $\mu$ ( ) $\mu$ (x) , x∊X

Definition 6.Let f: X ⟶ be the objective function, a fuzzy region (solution space) and S( ) the support of this region. The maximizing set over the fuzzy region, (f),is then define by its membership function [32].

μ ( )( )= { 0 if $f(x) \leq \inf_{S( )} f$ ; $\frac{( ) \quad ( )}{( ) \quad ( )}$ S( ) S( ) if $\inf_{S( )} f \leq f(x) \leq \sup_{S( )} f$ ; 1 if $\sup_{S( )} f \leq f(x)$ }

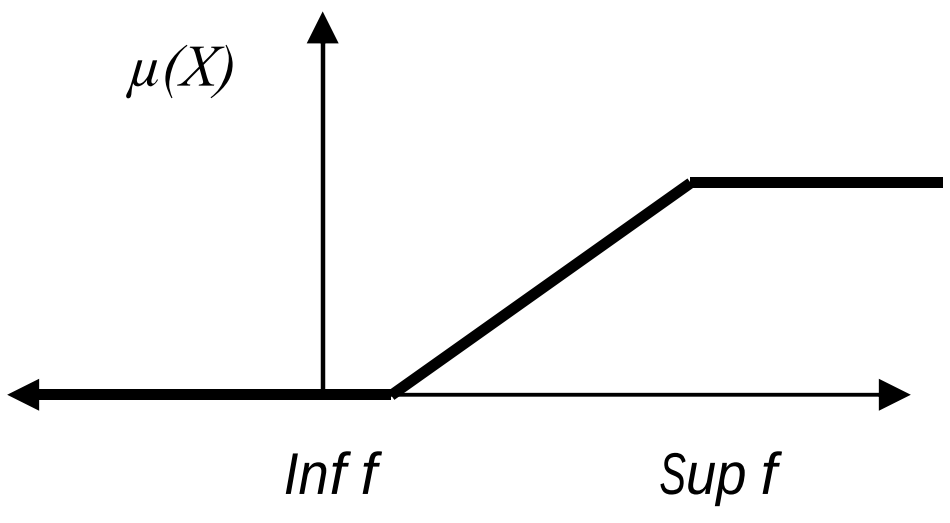

**Fig 1.** Fuzzy number

Definition 7.Suppose = ( , ) and = ( , , ) are triangular fuzzy numbers

So the arithmetic operations on them can be shown as [32]

⊕ =( + , + , + )

⊖ =( − , − , − )

⊗ =( ∗ , ∗ , ∗ )

⊘ ==( / , / , / )

## 3. Problem definition

In this paper a network of several nodes will be considered. Each node that could be considered as a customer has demand for service. The demand follows a time homogeneous Poisson process. Some servers are to be located at nodes of the network which means a subset of the nodes are to be chosen to locate one server in each. The service distribution is also Poisson and a maximum probability is considered for each server's occupancy. Each customer selects the server base on logit function of the distance and the objective is to maximize the covering of servers.

### 3.1. Mathematical formulation

The indices, parameters, independent parameters, decision variables, function and constraints are as follows:

**Indices:**

*i* and *j* are nodes number

**Parameters:**

: Distance between nodes *i* and *j*

: Demand rate for *i*th node (*i=1,2,3,…,n)*

: Obtained benefit rate in service center at node *j* from resourcing demand at node *i*

$\mu$ **:** The rate of Servicing in service centers number *j*

$\alpha$ : Probability of joining the queue line when the service center is occupied

$\beta$**:** Minimum probability of idleness at a service center in the long term

*MQl :* maximum queuing length

*n***:** Number of the network nodes (customers)

**Independent Parameters:**

:The recourse probability of the *i*th node demand to service center at the node *j*

**:** The occupancy probability (occupancy coefficient) of the service center at the *j*th

: The Length of queue

**Decision variables:**

*M*: Number of service centers

**Solution Answer:**

**:** Is the model decision variable and is1 if a server is located at node j and 0 otherwise

### 3.2. User choice

The system under study is represented as a network, where arcs are the possible paths between nodes, and nodes represent either candidate location for facilities or demand concentrations, or both. Most ‘‘central planning’’ location models assume that all the demand originating at a particular node is served by the same facility. This is not so in Competitive situations that competitors try to attract or capture as large a proportion of the demand as possible.

In this case, different percentages of the demand at each demand node will chose different facilities to patronize. The more attractive the facility for customers at a certain demand node, the percentage captures of the demand originating there will be larger. Furthermore, the percentage of customer capture by each facility will be given by Logit functions of the distance [32]. Hence, the probability of a user at node i choosing to go to the facility at node *j*, is defined by the expression:

$$=\frac{}{\sum} \qquad (1)$$

### 3.3. Probability of server idleness

This constraint refer to the probability of idleness of the service center, based on the fact that on the proposed model each demand responded with *(1-α)* and go to another center, so the idleness of each service center will be very important, so with using this constraint we will guarantee the probability of the idleness of the center in the value of β, and also in each service center it is possible to put only one center , and based on the exponential function of the service rate and demands are Poisson in each service center*(M/M/1)* it can be assumed. Now we should calculate the demand rate of each service center.
In order to $\varphi_i$ is demand rate for node*(i)* according to a Poisson process & node *(i)*with probability $p_{ij}$ go to server *(j)* thus we can calculate the rate of demand for each server by relation number (3) & we show it by $\bar{\varphi}_i$ .

$$=\sum \qquad \forall \qquad (2)$$

P (the probability of idleness for server *(j)*) $\geq \beta \quad \forall j \in n$ (3)

Now in order to in *M/M/1* the probability of idleness for server *(j)* gains by:

$_{0} = 1 - \frac{}{\mu}$ (4)

In order equation (5), equation (3) is:

$\sum \quad \leq \mu \, (1 - \quad )$ (5)

### 3.4. The objective function

In order relation (5) we can introduce the probability of idleness for each server and name it :

$= \frac{\sum}{\mu} \quad \forall j \in n$ (6)

Thus we can say that The occupancy probability for server *(j)* is and the idleness probability for server *(j)* is *(1- )*

referring to the definition of $B_{ij}$, the total benefit achieved at service centers would be equal to $\sum \; \sum$ if no demand is lost but in this objective function we assumed that the customer with especial probability wait in the queue. thus if each servers are idleness*(1- )* customers didn't wait in queue and we don't effect coefficient $\alpha$ in the objective function and if each servers are occupancy *(1- )* customer will wait with probability $\alpha$ in queue. Furthermore the objective function is:

$\sum \sum \quad ( \quad + \; 1 - \quad )$ (7)

In *M/M/1* expected number of customer waiting in the queue is calculated by:

$= \frac{}{( \quad )}$ (8)

Most of the times when a customer go to a service center if there is a queue in it, the most important factor that make a customer to go to the service center is the queue length, the probability of going to a service center by a customer will be reduced when the queue length is increased. In this model the probability of entering a customer to a queue will be a probability distribution and will be named as Maximum Queuing Length (MQL). MQL will be calculated based on the type of the service center and its last previous months or years records. considering as a linear probability function will make our model more similar to the real queuing models. In this model if the queue length is more than MQL the customer will not enter to the queue and =0 and if it will be less than MQL with the following probability it will be entered to the queue:

$= - \frac{}{} + 1$ (9)

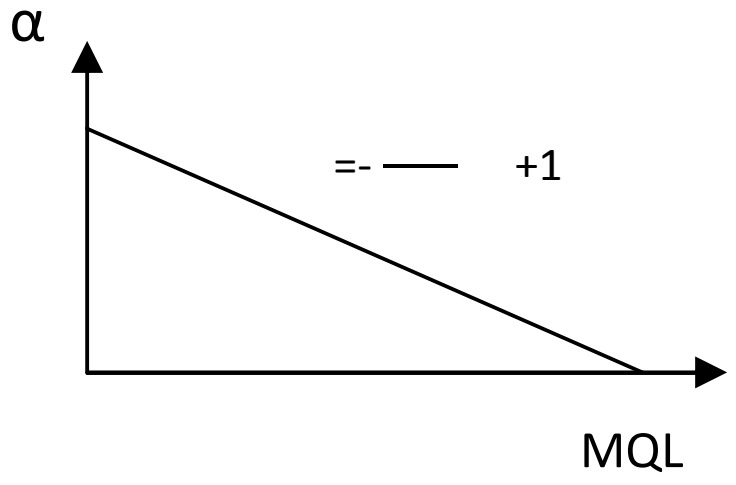

Fig 2. Probability of joining the queue line when the service center is occupied

Thus the objective function is:

$$\sum \sum \quad ( \quad - \quad + 1 \quad + \quad 1 - \quad ) \qquad (10)$$

For example if we want to establish some ATMs in a city to maximize the profit that is resulted from this decision. This objective function will be more important when this profit will be obtained based on the distance of the customers from service center. In other words, the profit of server j from the customer of node i ( )will be different from the customer of node k( ). the customer will choose the service center and the probability of this choosing is based on the distance of the customer to the node.

## 4. Mathematical modality (the Crisp Model)

We assume that all the nodes in the network are candidates to the location of facilities, as well as nodes containing demand. The entering firm wants to locate *M* facilities in the region. Note that the probabilities $p_{ij}$, which represent the customer–facility assignments, are a function of distance between node *( i )* and place *( j )*.objective is maximize the benefit of use customer from market in specific service center. Demand rate at each node is constraint.

$$\sum \sum \quad ( \quad - \quad + 1 \quad + \quad 1 - \quad ) \qquad (11)$$

$$= \frac{}{\sum} \qquad (12)$$

$$= \frac{\sum}{\mu} \qquad \forall j \in n \qquad (13)$$

$\sum \quad \leq \mu\,(1 - \quad) \quad \forall j \in n$ (14)

$\sum \quad = \quad \forall j \in n$ (15)

$\leq \quad \forall i, j \in n$ (16)

$\sum \quad = 1 \quad \forall i \in n$ (17)

$\in [0\ 1] \quad \forall i, j \in n$ (18)

$\in \{0, 1\} \quad \forall j \in n$ (19)

$( \ > 0, \beta > 0)$ (20)

The objective attempts to maximize the benefit of using customer from servers (11).The probability of a user at node i choosing to go to the facility at node *j* (12).The probability of occupancy for the service center at the *j*th node (13) .Constraint (14) ensures that the occupancy probability of each service center is not greater than $(1-\beta)$ **.**Constraint (15) specifics the number of facilities to be located by the entering firm. Constraint (16) forces customer capture only by open facilities. Although we use this constraint in the formal model, from a logical point of view this constraint is redundant, because the definition of the probabilities $p_{ij}$ forces these probabilities to be zero if there is no open facility at *j.* Thus, we do not use this constraint when solving the model. Constraint (17) ensures that a 100% of customers at a demand node i will be served somewhere. Again, this constraint is not needed because of the definition of the probabilities $p_{ij}$ ensures that the summation of them will be exactly 1. However, it is stated only for the reason of making the model more understandable. Finally, constraints (18)–(20) ensure non-negativity, integrality and bounds on the variables.

## 5. The fuzzy queuing maximal benefit location model (FQMBLM)

This section is devoted to the definition of the stages of building the model. The model's parameters and variables, as well as the fuzzy sets, constraints and the objective function will be presented.

### 5.1. The parameters and decision variables

The following is a list of the parameters used in the model:
M: number of service centers (a crisp number)
$p_{ij}$*:* the recourse probability of the *i*th node demand to service center at the node *j*.(a crisp number)
: Distance between nodes *i* and *j*. (a crisp number)
~ ( , , ) :the demand rate for service at node i (a triangular fuzzy number)
~( , , ):the service rate of server j (a triangular fuzzy number)

~( , , ):the Minimum probability of idleness at a service center in the long term(a triangular fuzzy number).

~( , , ): The occupancy probability (occupancy coefficient) of the service center at the *j*th.(a triangular fuzzy number).

~ ( , , ): the length of queue (a triangular fuzzy number).

~ ( , , ): Probability of joining the queue line when the service center is occupied(a triangular fuzzy number).

the predefined coefficient for the truth value of fuzzy queue constraint (a number between zero and one).

Is 0–1 variable which assumes value 1 if node i is covered by server j, and 0 otherwise

Is 0–1 variable; it turns 1 if a server is located at node j and 0 otherwise

### 5.2. The fuzzy sets and proposed the objective function

In this section we first define the model's fuzzy sets and then introduce the objective function. As far as our model is a non-linear model, changing this model to the nonlinear fuzzy model will be as follows:

1) Making the Objective function fuzzy
2) Making the constraints fuzzy

Our goal is maximizing the triangular fuzzy values. So we can maximize the center and right side and minimize left side.

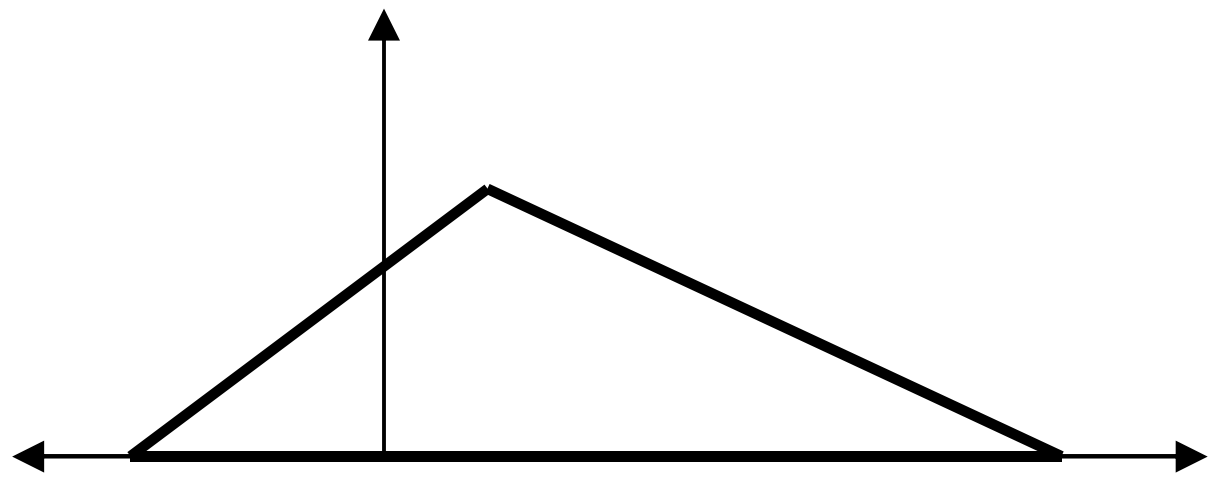

**Fig 3.** A triangular fuzzy number

Using aforementioned method will change our model to the multi-objective problem as follows:

.

Minimize = ( - )X (20)
Maximize = ( )X (21)
Maximize = ( − )X (22)

S.t:

$$\begin{cases} AX \leq B \\ X \geq 0 \end{cases} \quad (23)$$

A method for solving the problem is using dependency functions for three objective functions and of them will be maximize, so:

$$= MIN( \quad - \quad ))X, \quad = MAX( \quad - \quad )X \quad (24)$$

$X \in x$ $\qquad$ $X \in x$

$$= MAX( \quad )X, \quad = MIN( \quad )X \quad (25)$$

$X \in x$ $\qquad$ $X \in x$

$$= MAX( \quad - \quad )X, \quad = MIN( \quad - \quad )X \quad (26)$$

$X \in x$ $\qquad$ $X \in x$

, and will be as follows:

$$=MIN\sum [(\sum \quad )((1- \quad )+(- — \quad +1)( \quad )) -\sum [(\sum \quad )((1- \quad )+(- — \quad +1)( \quad ))$$

$x \epsilon X$ (27)

$$=MAX\sum [(\sum \quad )((1- \quad )+(- — \quad +1)( \quad ))$$

$x \epsilon X$ (28)

$$=MAX\sum [(\sum \quad )((1- \quad )+(- — \quad +1)( \quad )) -\sum [(\sum \quad )((1- \quad )+(- — \quad +1)( \quad ))$$

$x \epsilon X$ (29)

$X=\{X|AX \leq b, X \geq 0\}$ and the results of will be named as maximize desirable results and as minimize desirable results. We will use following dependency functions for describing objective functions as follows:

$$\mu = \begin{cases} 1 & ( \quad - \quad )X \leq \\ \frac{(( \quad - \quad )X)}{} & \leq ( \quad - \quad )X \leq \end{cases} \quad (30)$$

$$0 \qquad ( \quad - \quad )X \geq$$

$$\mu \; = \begin{cases} 1 & ( \quad )X > \\ \dfrac{(( \qquad ) \; )}{} & \leq \; ( \quad )X \leq \\ 0 & ( \quad )X < \end{cases} \tag{31}$$

$$\mu \; = \begin{cases} 1 & ( \quad - \quad )X > \\ \dfrac{(( \qquad ) \; )}{} & \leq \; ( \quad - \quad )X \leq \\ 0 & ( \quad - \quad )X < \end{cases} \tag{32}$$

Finally we will solve following standard programming:

$$\begin{cases} \text{Max} ( \; ) \\ \boldsymbol{\mu} \;\; \geq \;\; , \boldsymbol{\mu} \;\; \geq \;\;\; , \boldsymbol{\mu} \;\; \geq \\ Ax \leq b \\ x \geq 0 \end{cases} \tag{33}$$

Making the constraints fuzzy

**Lemma1.**Given two triangular fuzzy numbers $= ( \; , \; , \; )$ $= ( \; , \; , \; )$, we have [34]:

$$T ( \;\; \leq \;\; ) = 1 \Longleftrightarrow \quad \leq \quad , \tag{34 }$$

$$T ( \;\; \leq \;\; ) \leq 1 \Longleftrightarrow \quad \leq \quad - \qquad ( \quad - \quad ), \tag{35 }$$

The constraint becomes the following:

$$T \left( \frac{\Sigma}{} \leq (1- \;\; ) \right) \geq \tag{34}$$

$$\frac{\Sigma}{\quad} \leq (1- \quad ) - \quad ((1 - \quad ) - (1 - \quad )) \quad (35)$$

The FQMBLM finally is transformed to a 0_1 integer programming model as:

$$\text{Max} ( \ ) \quad (36)$$

$$\boldsymbol{\mu} \ \geq \quad (37)$$

$$\boldsymbol{\mu} \ \geq \quad (38)$$

$$\boldsymbol{\mu} \ \geq \quad (39)$$

$$= \frac{\quad}{\Sigma} \quad (40)$$

$$\frac{\Sigma}{\mu} \leq (1- \quad ) - \quad ((1 - \quad ) - (1 - \quad )) \quad (41)$$

$$\sum \quad = \quad \forall j \in n \quad (42)$$

$$\leq \quad \forall i, j \in n \quad (43)$$

$$\sum \quad = 1 \quad \forall i \in n \quad (44)$$

$$\in [0\ 1] \quad \forall i, j \in n \quad (45)$$

$$\in \{0, 1\} \quad \forall j \in n \quad (46)$$

## 6. Solution Methodology

In this section we provide a solution methodology for proposed model. Since the *p*-median problem is NP-Hard [35], and the derived 0–1 integer programming model in this problem can be reduced to the p-median problem in polynomial time, so it is NP-Hard. Therefore it cannot be solved in general to optimality and it is appropriate to develop a heuristic method to solve the problem within a reasonable time [34]. To this aim, Two Heuristic algorithms are developed to solve the problem.

We run each algorithm 7 times that is contain the results of maximum desirable results and minimum desirable results .

*Step0.Calculation of*

*Step1. Calculation of*

*Step2. Calculation of*

*Step3.*

*Step4.*

*Step5.*

*Step6.Fuzzy Objective Function*

### 6.1. Proposed a new Genetic Algorithm

In this paper we will use Alp et al. [34] proposed Genetic Algorithm, because of its simple and fast method in solving problems and its excellent capability to generate solutions. The algorithm will be discussed as follows, for more explanation on this algorithm Alp et al. Research is proposed.

#### 6.1.1. Encoding and Fitness Function

We will use a simple encoding where the genes of a chromosome correspond to the indices of the selected facilities. The fitness function will be easily calculated with using the problem data.

#### 6.1.2. Population Size and initializing the population

The two important factors of population size as: every gene must be present in the initial problem and the population size should be proportional to the number of solutions will be responding to extension of feasible solutions with formula proposed in follow.

*Population Size(total nodes, number of servers)=* $\dfrac{\;,\;\frac{\quad}{\quad}*\frac{(\;)}{\quad}\;*}{\quad}$ (47)

Let S=C be the number of all possible solutions to problem, and

d= $\frac{\quad}{\quad}$ the rounded-up density of the problem.

If ———— is an integer then each gene is represented in the initial population with an equal frequency. If ————is not an integer then after disturbing all of the genes from *1* to *number of servers* to each group, we allocated random genes to fill empty slots.

### 6.1.3. Generating new members

Different with previous algorithm, we use Alp et al [36] proposed method for generating new members. They take the union of the genes of the parents, obtaining an infeasible solution with *m* genes where *m>total nodes* and then for reducing the number of genes by one, discard the genes whose discarding produces the best fitness function value until reach *total nodes*. However, genes that are present in both parents never must be dropped. We call the infeasible solution obtained after the union operation the "*draft member*" and the feasible solution generated by the heuristic as the "*candidate member*". The input of the generation process is two different members and the output will be a candidate member.

### 6.1.4. Mutation and Replacement

As far as the mutation operator is negligible we decided not to use it. The replacement operator will be operated only on $N'$ . The steps for the replacement operator are as follows [28]:

*Input:* One candidate member.

*Step 1.* If fitness value of the input candidate member is higher than the maximum fitness value in the population, then discard this candidate member and terminate this operator.

*Step 2.* If the candidate member is identical to an existing member of the current population, then discard this candidate member and terminate this operator.

*Step 3.* Replace the worst member of the population with the input candidate member.

*Step 4.* Update the worst member of the population.

*Step 5.* Update the best member of the population.

*Output:* Population after introducing the candidate member.

### 6.1.5. Termination

The algorithm terminates after observing $\lfloor total\,nodes\sqrt{number\ of\ servers}\rfloor$ successive iterations reach the best solutions and after observing $\lfloor total\,nodes\sqrt{number\ of\ servers}\rfloor^2$ successive iterations where the best solution found has not changed. The iteration consists of one use of the generation and replacement operators.

### 6.2. Proposed a new Ant Colony Optimization:

The principle of ACO algorithms [37], [38] is based on the way ants search for food. Each ant takes into consideration (probabilistic choice) pheromone trails left by all other ant colony members which preceded its course, the pheromone trail being a trace, a smell left by every ant on its way. This pheromone evaporates with time, and therefore the probabilistic choice for each ant changes with time. After many ant courses, the path to the food will be characterized by higher pheromone traces and thus all ants will follow the same path. This collective behavior, based upon a shared memory among all colony ants could be adapted and used for solving combinatorial optimization problems with the following analogies:

The real ant search space becomes the space of the combinatorial problem solutions. The amount of food inside a source becomes the evaluation of the objective function for the corresponding solution. The pheromone trails become an adaptive shared memory.

In the following subsections, we discuss the proposed ACO algorithm in detail, the generation of the initial solution, the calculation of the objective function value and the parameters that are used.

#### 6.2.1. Propose an initiative index:

In this section we propose an initiative index related to fuzzy queuing location problem. This problem has to important factor for customers and decision maker. These factors are distance and service rate. Customers who's want to use these servers want in minimum time get service from server. Thus rate of servicing is really important for decision maker to locate servers in special node and the least distance is really important for customers.

For each node of j we will submit all distance of it from other nodes, so less d (potentially les distance from the node j) and more φ (mores service rate in a service center) means the importance of a node and a server. So desired allocation of the server i to the node j will be as follows:

$$\begin{cases} \propto \\ \Longrightarrow \quad \propto \; . - \\ \propto - \end{cases} \tag{48}$$

Our main motivation to use this criterion is ant can find the best solutions directly and servers are located in the locations with the minimum lenght. The value of this criterion is constant in each repeat.

**6.2.2. Selection probability:**

An ant *i* choose node *j* to assign to location 1 by the following probability [39]:

$$(\;) = \frac{[\;(\;)]\;\;[\;\;]}{\Sigma \quad [\;(\;)]\;\;[\;\;]} \tag{49}$$

( ) is the pheromone impact in node j at repeat t, and , are parameters that will result us to find pheromone impact and meta-heuristic criterion. Base on our model some changes effected on this equation. is contained nodes that ant i does not put a server on node j and still is empty.

$\Sigma$ =1 and this step is repeat to allocate all servers to nodes by each ants.

**6.2.3. Update Pheromone:**

Updating pheromone is based on the following equation [39]:

$$(\; + 1) = \; . \;\; (t) + \Sigma \tag{50}$$

$\rho$ is persistence of pheromone(its value is $0<\rho<1$) such that $(1 - \rho)$ represents the percentage of evaporation trail between time $t$ and $t+n$ and it helps to algorithm to forgive worth answers. Where is the quantity per unit of trail substance (pheromone in real ant) laid on node *j* by *kth* ant and base on maximize or minimize of objective function is define.

ماندگاری اثر فرمون و مقدار آن 1> 0< است، در نتیجه ( 1- ) در صد تبخیر فرمون را بیان می کند.پارامتر به دلیل پرهیز از انباشتن شدن بیش از حد فرمون در یک یال استفاده می شود و اجازه می دهد که الگوریتم انتخاب های بد انجام شده را به فراموشی بسپارد مقدار فرمونی است که مورچه k در گره *j* ام می گذارد و برحسب ماکسیموم سازی و مینیمم ساز بودن تابع هدف به صورت زیر تعریف می شود:

$$( \qquad )= \begin{cases} \Theta.\frac{}{\text{Ƒ}} & \textit{If the and k allocate a server on node j} \\ 0 & \textit{Otherwise} \end{cases} \qquad (51)$$

Ƒجواب تابع هدف مورچه kام می باشدو چون هدف مینیم سازی می باشد هر چه مقدار تابع هدف کوچکتر شود مقدار کسر Θ /Ƒ بیشتر می باشد و باعث افزایش احتمال انتخاب این گره توسط مورچه های دیگر می باشد و همچنین Θ حداکثر مقدار فرمونی است که می تواند از یک مورچه در یک یال باقی بماند.

Ƒ Is the objective function of ant's number *kth*.if our objective function is minimize, by minimizing objective function the amount of Θ /Ƒ is maximized and the probability of choosing this nodes by other ants is increased.Θ is the quantity of maximum pheromone in each node.

When the objective function is maximizing:

$$( \qquad )= \begin{cases} \Theta.\text{Ƒ} & \textit{If the and k allocate a server on node j} \\ 0 & \textit{Otherwise} \end{cases} \qquad (52)$$

### 6.2.4. Ant Population size:

The two important factors of population size as: every solution algorithm must be present in the initial problem and the population size should be proportional to the number of solutions will be responding to extension of feasible solutions with formula proposed in follow.

*Ant Population Size (total nodes, number of servers) =*

$$* \max\ 2, \quad * \frac{()}{} \quad * \tag{53}$$

Let s=c is the number of all possible solutions to problem, and the rounded-up density of the problem.

### 6.2.5Termination:

The algorithm terminates after observing * successive iterations reach the best solutions and after observing * − successive iterations where the best solution found has not changed. The iteration consists of one use of the generation and replacement operators.

## 7. Numerical Examples and the Results:

To solve the problems, a MATLAB R 2008a computer program was used to obtain the local optimum solution of the same problems with GA and Ant Colony Optimization algorithms. We use JMP version 8.0 to setting algorithm parameters. At first we run a 20nodes problem that (Table 1) Fuzzy Demand rates of the nodes ( , , ), distances between the nodes ( ) and service rate: ( , , )are randomly generated from([104, 181] ,[54,131] ,[4,80]) , [6, 25] and ([244 ,290],[194,240],[144,190]) respectively. Other problem parameters including Minimum probability of idleness at a service center in the long term ( , , ) ,Maximum queeing length $MQl$, Arrival rate and are assumed (0.1 ,0.15 ,0.2 ) , 25, 100 and 0.5.

Ant Colony Optimization parameters are *Evaporation Rate, Maximum pheromone, Population coefficient,* , . Ant colony algorithm has 5 parameters, therefore maximum number of experiment for this especial problem is2 .

As far as the GA algorithm doesn't have any parameters the GA parameters are opted base on problem parameters.

The amount of Ant Colony Optimization algorithm parameters after setting parameters will be as following: *Evaporation Rate=0.97, Maximum pheromone=200, Population coefficient=2, =0.75,* = 0.75. In table 3 distinctions of these two algorithms (ACO and GA) with set parameters will be considered.

The same process had been done for problems with 30, 40, 50, 60 node. But because of the limited space the results will be only considered for the 30 to 70-node problems. In Tables 4–7 a comparison between results based on the Genetic Algorithm and Ant Colony Optimization algorithm is drawn and the percent of the difference amongst these algorithm is represent. A computer with Dual core 2.0 GHz CPU, 1.00 Cache and 3.0 GB RAM was used. Finally in the figure 1 and 2 a comparison of these two solution algorithms based on the running time and objective function will be considered.

## 8. Conclusions and future research:

In this paper a mathematical location model to maximize service profit is considered. Queuing theory and fuzzy arrangement for making the problem more realistic was used and this model is changed to integer zero and one programming. The proposed model is in NP-hard category and is extended for Genetic algorithm and Ant Colony. The proposed Genetic Algorithm has better results with longer run time.

Changing the model to multi-objective models, changing model to multi-objective fuzzy model and improving the proposed ant colony can be considered as an extension to our research.

Table1. Demand, service rate and distance for 20 nodes

| | 1 | 2 | 3 | 4 | 5 | 6 | 7 | 8 | 9 | 10 | 11 | 12 | 13 | 14 | 15 | 16 | 17 | 18 | 19 | 20 |
|---|---|---|---|---|---|---|---|---|---|---|---|---|---|---|---|---|---|---|---|---|
| | 160 | 179 | 173 | 145 | 181 | 143 | 104 | 145 | 159 | 181 | 164 | 147 | 112 | 149 | 150 | 112 | 143 | 166 | 137 | 180 |
| | 110 | 129 | 123 | 95 | 131 | 93 | 54 | 95 | 109 | 131 | 114 | 97 | 62 | 99 | 100 | 62 | 93 | 116 | 87 | 130 |
| | 60 | 79 | 73 | 45 | 81 | 43 | 4 | 45 | 59 | 81 | 64 | 47 | 12 | 49 | 50 | 12 | 43 | 66 | 37 | 80 |
| | | | | | | | | | | | | | | | | | | | | |
| | 289 | 273 | 263 | 254 | 245 | 271 | 277 | 271 | 275 | 250 | 251 | 259 | 244 | 290 | 261 | 275 | 270 | 275 | 276 | 272 |
| | 239 | 223 | 213 | 204 | 195 | 221 | 227 | 221 | 225 | 200 | 201 | 209 | 194 | 240 | 211 | 225 | 220 | 225 | 226 | 222 |
| | 189 | 173 | 163 | 154 | 145 | 171 | 177 | 171 | 175 | 150 | 151 | 159 | 144 | 190 | 161 | 175 | 170 | 175 | 176 | 172 |
| | | | | | | | | | | | | | | | | | | | | |
| 1 | 0 | 2 | 33 | 22 | 18 | 32 | 31 | 11 | 8 | 18 | 30 | 32 | 33 | 7 | 32 | 17 | 20 | 28 | 21 | 27 |
| 2 | | 0 | 4 | 28 | 22 | 23 | 6 | 10 | 34 | 26 | 30 | 35 | 29 | 35 | 31 | 30 | 30 | 7 | 26 | 29 |
| 3 | | | 0 | 21 | 10 | 14 | 34 | 4 | 5 | 2 | 30 | 24 | 14 | 17 | 23 | 8 | 17 | 35 | 21 | 29 |
| 4 | | | | 0 | 35 | 28 | 22 | 27 | 1 | 17 | 9 | 17 | 31 | 19 | 35 | 24 | 16 | 33 | 28 | 28 |
| 5 | | | | | 0 | 31 | 29 | 29 | 2 | 17 | 30 | 2 | 16 | 6 | 3 | 6 | 1 | 12 | 27 | 21 |
| 6 | | | | | | 0 | 9 | 7 | 33 | 21 | 32 | 5 | 9 | 31 | 15 | 11 | 2 | 24 | 12 | 34 |
| 7 | | | | | | | 0 | 6 | 1 | 31 | 33 | 21 | 20 | 7 | 33 | 24 | 18 | 25 | 23 | 10 |
| 8 | | | | | | | | 0 | 26 | 1 | 13 | 34 | 21 | 33 | 29 | 18 | 30 | 2 | 17 | 22 |
| 9 | | | | | | | | | 0 | 23 | 24 | 33 | 15 | 11 | 28 | 26 | 12 | 27 | 1 | 30 |
| 10 | | | | | | | | | | 0 | 4 | 6 | 27 | 5 | 9 | 17 | 21 | 14 | 11 | 13 |
| 11 | | | | | | | | | | | 0 | 7 | 21 | 5 | 24 | 29 | 30 | 17 | 8 | 18 |
| 12 | | | | | | | | | | | | 0 | 19 | 10 | 10 | 22 | 22 | 1 | 31 | 3 |
| 13 | | | | | | | | | | | | | 0 | 31 | 22 | 32 | 2 | 24 | 30 | 16 |
| 14 | | | | | | | | | | | | | | 0 | 6 | 29 | 34 | 13 | 2 | 22 |
| 15 | | | | | | | | | | | | | | | 0 | 22 | 25 | 16 | 23 | 3 |
| 16 | | | | | | | | | | | | | | | | 0 | 29 | 6 | 25 | 32 |
| 17 | | | | | | | | | | | | | | | | | 0 | 30 | 3 | 19 |
| 18 | | | | | | | | | | | | | | | | | | 0 | 18 | 22 |
| 19 | | | | | | | | | | | | | | | | | | | 0 | 31 |
| 20 | | | | | | | | | | | | | | | | | | | | 0 |

Table 2. The proposed Ant colony Optimization algorithm Results for 20 nodes

| | Evaporation rate | Maximum Pheromone | Population Coefficient | | | Objective Function |
|---|---|---|---|---|---|---|
| 1 | 0.95 | 150 | 1 | 0.5 | 0.5 | 0.916260417202333 |
| 2 | 0.95 | 150 | 1 | 0.5 | 1 | 0.890040041689561 |
| 3 | 0.95 | 150 | 1 | 1 | 0.5 | 0.924774503406953 |
| 4 | 0.95 | 150 | 1 | 1 | 1 | 0.847257131688177 |
| 5 | 0.95 | 150 | 3 | 0.5 | 0.5 | 0.942346372892224 |
| 6 | 0.95 | 150 | 3 | 0.5 | 1 | 0.961599078765475 |
| 7 | 0.95 | 150 | 3 | 1 | 0.5 | 0.945567230748633 |
| 8 | 0.95 | 150 | 3 | 1 | 1 | 0.900613133637894 |
| 9 | 0.95 | 250 | 1 | 0.5 | 0.5 | 0.942429086997071 |
| 10 | 0.95 | 250 | 1 | 0.5 | 1 | 0.903482754282721 |
| 11 | 0.95 | 250 | 1 | 1 | 0.5 | 0.917646318024575 |
| 12 | 0.95 | 250 | 1 | 1 | 1 | 0.908057738900665 |
| 13 | 0.95 | 250 | 3 | 0.5 | 0.5 | 0.959454852517375 |
| 14 | 0.95 | 250 | 3 | 0.5 | 1 | 0.926837710481436 |
| 15 | 0.95 | 250 | 3 | 1 | 0.5 | 0.886847013133750 |
| 16 | 0.95 | 250 | 3 | 1 | 1 | 0.921018556383610 |
| 17 | 0.99 | 150 | 1 | 0.5 | 0.5 | 0.831600601615242 |
| 18 | 0.99 | 150 | 1 | 0.5 | 1 | 0.833925490928573 |
| 19 | 0.99 | 150 | 1 | 1 | 0.5 | 0.894126378300047 |
| 20 | 0.99 | 150 | 1 | 1 | 1 | 0.967775195459996 |
| 21 | 0.99 | 150 | 3 | 0.5 | 0.5 | 0.709922919636424 |
| 22 | 0.99 | 150 | 3 | 0.5 | 1 | 0.740412882743890 |
| 23 | 0.99 | 150 | 3 | 1 | 0.5 | 0.863524377090585 |
| 24 | 0.99 | 150 | 3 | 1 | 1 | 0.874160581795046 |
| 25 | 0.99 | 250 | 1 | 0.5 | 0.5 | 0.758405214335774 |
| 26 | 0.99 | 250 | 1 | 0.5 | 1 | 0.700803878389285 |
| 27 | 0.99 | 250 | 1 | 1 | 0.5 | 0.835743756941273 |
| 28 | 0.99 | 250 | 1 | 1 | 1 | 0.844284103035315 |
| 29 | 0.99 | 250 | 3 | 0.5 | 0.5 | 0.594087944877577 |
| 30 | 0.99 | 250 | 3 | 0.5 | 1 | 0.770083843348551 |
| 31 | 0.99 | 250 | 3 | 1 | 0.5 | 0.894155189229175 |
| 32 | 0.99 | 250 | 3 | 1 | 1 | 0.831769609306284 |

Table 3. A comparison of the results obtained from the GA against ACO for 20 nodes problem

| **GA** | | | **ACO** | | | **%** |
|---|---|---|---|---|---|---|
| Objective | Run time | Cover nodes | Objective | Run time | Cover nodes | 2.5 |
| 0.975205178 | 27 | 1,3,18,19,20 | 0.950122509 | 9 | 9,15,16,19,20 | |

Table 4. A comparison of the results obtained from the GA against ACO for 30 nodes problem

| **GA** | | | **ACO** | | | **%** |
|---|---|---|---|---|---|---|
| Objective | Run time | Cover nodes | Objective | Run time | Cover nodes | 5.4 |
| 0.957831127 | 171 | 1,6,20,26,29 | 0.905217011 | 98 | 1,13,14,15,24 | |

Table 5. A comparison of the results obtained from the GA against ACO for 40 nodes problem

| | GA | | | ACO | | % |
|---|---|---|---|---|---|---|
| Objective | Run time | Cover nodes | Objective | Run time | Cover nodes | 4.3 |
| 0.950901828 | 739 | 1,28,33,34,37 | 0.910000207 | 577 | 14,20,32,36,37 | |

Table 6. A comparison of the results obtained from the GA against ACO for 50 nodes problem

| | GA | | | ACO | | % |
|---|---|---|---|---|---|---|
| Objective | Run time | Cover nodes | Objective | Run time | Cover nodes | 3.4 |
| 0.960653412 | 2501 | 7,26,34,41,45 | 0.927938959 | 1277 | 9,23,27,31,39 | |

Table 7. A comparison of the results obtained from the GA against ACO for 60 nodes problem

| | GA | | | ACO | | % |
|---|---|---|---|---|---|---|
| Objective | Run time | Cover nodes | Objective | Run time | Cover nodes | 7.0 |
| 0.972728679 | 6433 | 24,32,39,41,57 | 0.904487024 | 3235 | 15,2934,39,57 | |

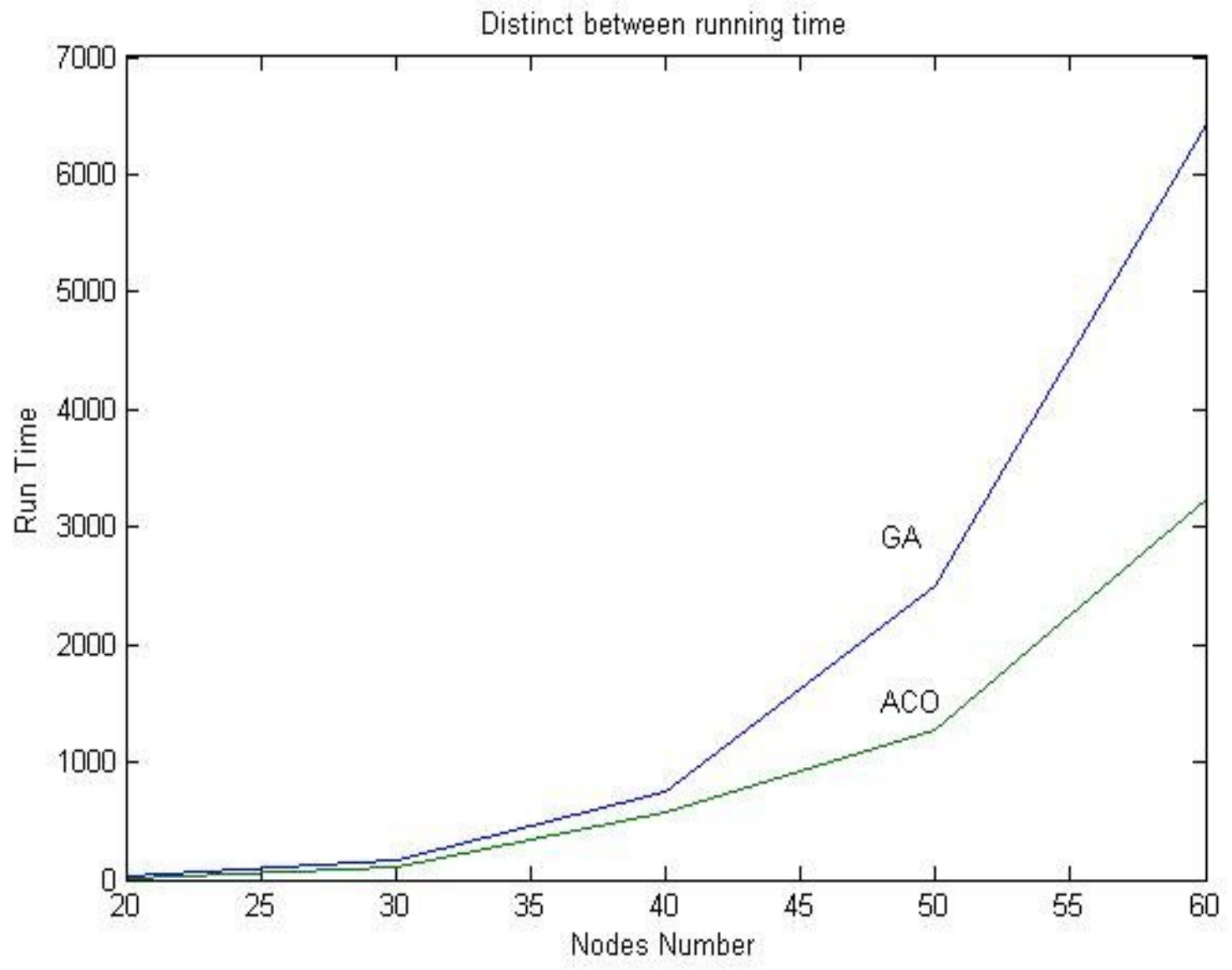

**Fig4.** Comparison of GA and ACO Based on the Run time Criterion

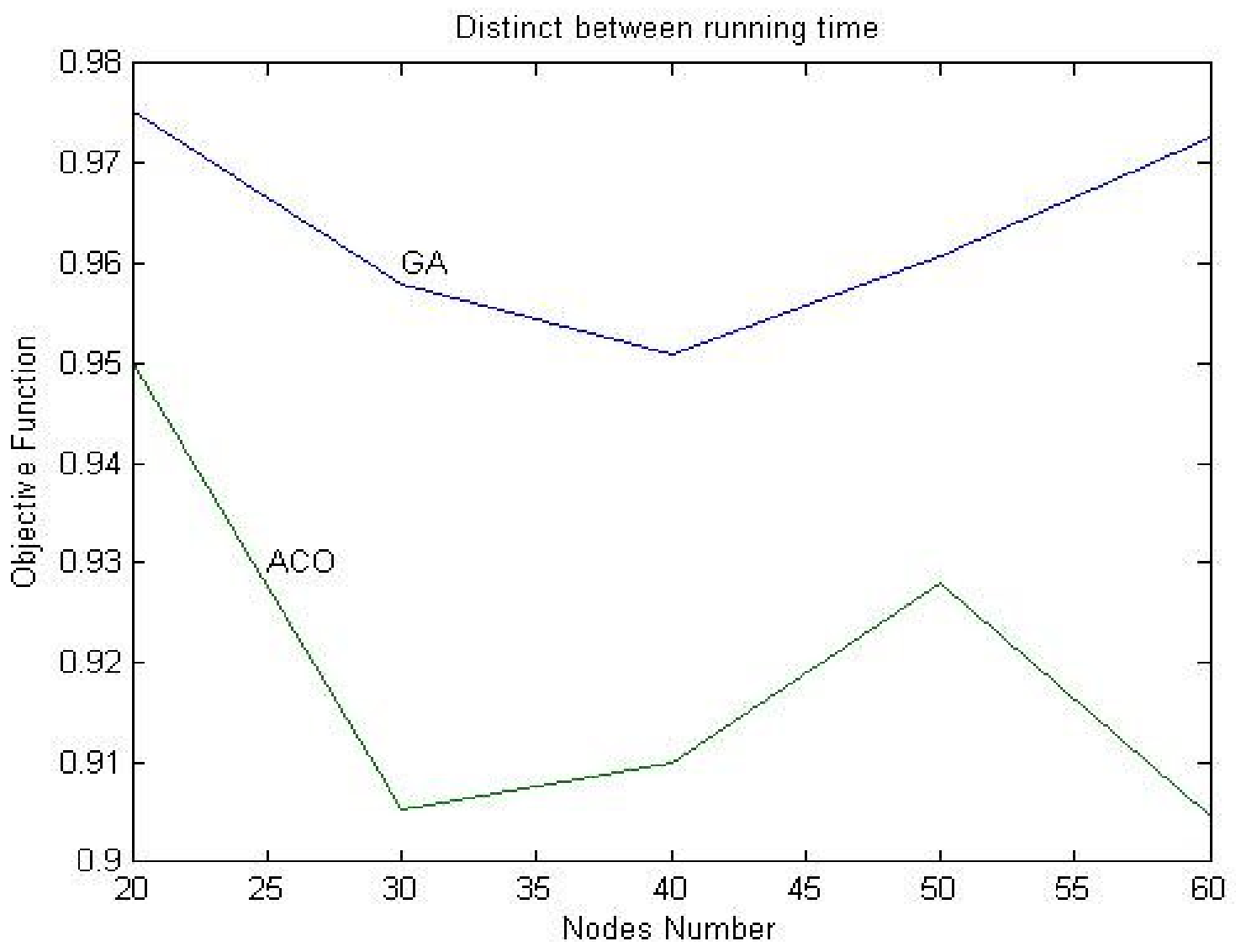

**Fig 5.** Comparison of GA and ACO Based on the Objective Function Criterion